\documentclass[reprint,aps]{revtex4-1}

\usepackage{amsmath,latexsym,xcolor,mathtools}
\usepackage{graphicx,verbatim, subcaption, caption}
\usepackage{color}
\DeclareMathOperator{\Tr}{Tr}
\usepackage[title]{appendix}

\def\bra#1{\mathinner{\langle{#1}|}}
\def\ket#1{\mathinner{|{#1}\rangle}}
\def\braket#1{\mathinner{\langle{#1}\rangle}}

\begin{document}
\title{An Optomechanical Quantum Cavendish Experiment}

	\author{Abdulrahim Al Balushi}
	\affiliation{Department of Physics and Astronomy, University of Waterloo, Waterloo, Ontario, N2L 3G1, Canada}
         \author{Wan Cong}
	\affiliation{Department of Physics and Astronomy, University of Waterloo, Waterloo, Ontario, N2L 3G1, Canada}
	\affiliation{Institute for Quantum Computing, University of Waterloo, Waterloo, Ontario, N2L 3G1, Canada}
	\author{Robert B. Mann}
	\affiliation{Department of Physics and Astronomy, University of Waterloo, Waterloo, Ontario, N2L 3G1, Canada}
	\affiliation{Institute for Quantum Computing, University of Waterloo, Waterloo, Ontario, N2L 3G1, Canada}

\date{\today}
\begin{abstract}
An open question in experimental physics is the characterization of gravitational effects in quantum regimes. 
We propose an experimental set-up that uses well-tested techniques in cavity optomechanics to observe the effects of the gravitational interaction between two micro-mechanical oscillators on the interference of the cavity photons through the shifts in the visibility of interfering photons. The  gravitational coupling leads to a shift in the  period and magnitude of the visibility whose observability is within reach of current technology.  We discuss the feasibility of the set-up as well as 
the effects on entanglement due to gravitational interaction.
\end{abstract}

\pacs{Valid PACS appear here}
\maketitle

\section{Introduction}
One of the biggest difficulties in constructing a theory of quantum gravity is the lack of experimental data. Unavailability of clean data from regimes where both quantum and gravitational effects are present have cast a long shadow on the fundamental conceptual problems that a theory of quantum gravity is expected to solve \cite{Kiefer2006,Woodard2009}. Although both theories have been successfully tested to extremely high degrees in their respective domains of validity, the disparities between them (i.e. large distances and massive bodies for general relativity versus short distances and small masses for quantum mechanics), which stem from the weakness of gravity and the decoherence of quantum states, have led to the yet-unsurmounted task of designing experiments that can access regimes where both theories predict effects of comparable degrees of observability. 

These experiments are of two types: 1) those where the goal is only to construct a measurement apparatus sensitive enough to provide information about cosmological and astrophysical phenomena, or 2) those experiments where both the source of observations and the measurement apparatus need to be constructed. The former include observations of the primordial cosmic microwave background (CMB) for information about the very early universe (i.e. a rare example of a natural quantum gravity regime), and sensitive detection of gravitational waves from black holes mergers as a possible source of information about the quantum degrees of freedom inside black holes \cite{Barcelo2017}. The latter approach was first proposed by Feynman  \cite{Feynman1957}, where he suggested putting a massive object in superposition to test whether its gravitational field can also be put in superposition (i.e. is quantum in nature) or whether a "gravitational collapse" would prevent this from happening. 

Advances in optomechanics \cite{Aspelmeyer2014} and atom interferometry \cite{Asenbaum2017} have made the possibility of measuring the effects of gravity in table-top quantum systems closer than ever. Another promising route exploits advanced satellite technologies that will allow quantum protocols to be tested over large length scales where the effects of gravity and spacetime curvature are expected to be non-trivial \cite{Rideout2012}. 

In this paper we investigate the question: given a model of gravitational interaction between two quantized systems, how can we experimentally observe the effects of this interaction? To this end we propose an optomechanical set-up to observe the effect of the gravitational interaction between two quantum micro-mechanical oscillators. Such a set-up  involving superposing mirrors of order $10^{14}$ atoms was proposed in \cite{Marshall2003}, and its application in observing the effects of gravitational decoherence models was considered in \cite{Adler2005}. Here we assume that the gravitational interaction is Newtonian gravity $GMm/|\hat{r}_1-\hat{r}_2|$, where $\hat{r}_1$ and $\hat{r}_2$ are position operators of the gravitating masses,  and calculate its effect  on the visibility pattern of interfering photons in an optomechanical set-up perturbatively. We find that the gravitational coupling leads to an observable shift in the period and magnitude of the visibility of photons that is within reach of today's technologies. 

Observation of the effects of models of gravity that modifies quantum mechanics, such as gravitational decoherence and semi-classical gravity, in optomechanical settings has been considered before \cite{Chen2013}. Here, we add to this list of signatures the one that we calculate from the Newtonian quantum gravity model, $GMm/|\hat{r}_1-\hat{r}_2|$. 

Our paper is organized as follows. We first discuss the set-up to be used to search for the model's signatures, and the parameters that will optimize between their strength and experimental feasibility. The nature and magnitude of the signatures is then discussed, as well as the requirements to deal with environmental decoherence. We sum up our results in a concluding section.
\section{Experimental proposal}
\begin{figure*}[htb]
 \centering
    \begin{subfigure}{0.40\textwidth}
        \includegraphics[width=1.0\linewidth]{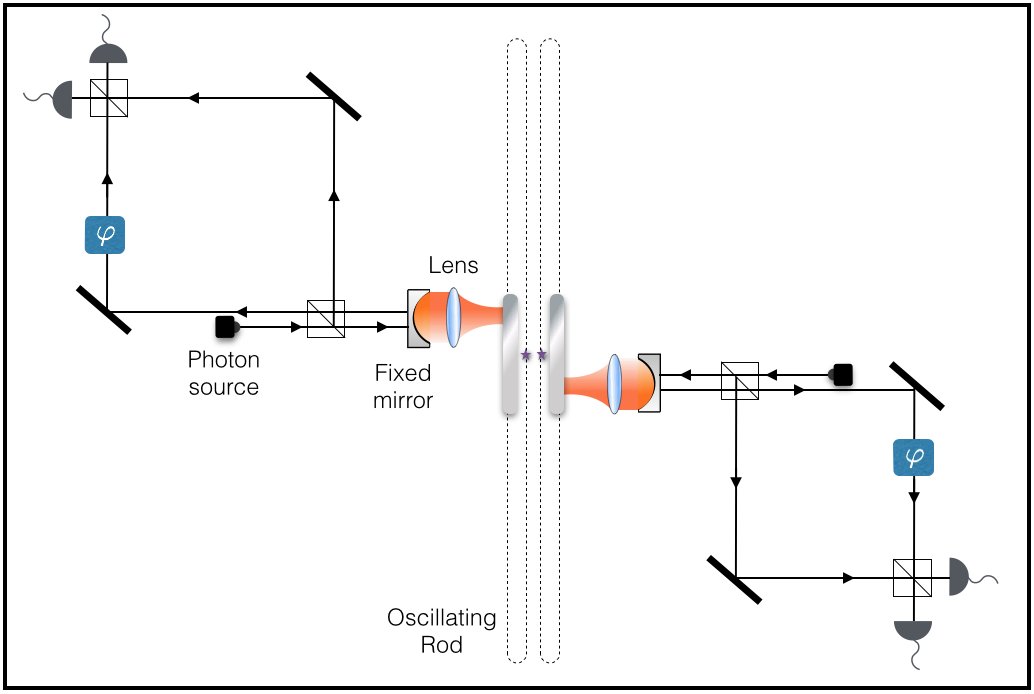}
        \caption{upper-view of the set-up}
        \label{fig:p1}
    \end{subfigure}
    \begin{subfigure}{0.40\textwidth}
        \includegraphics[width=0.92\linewidth]{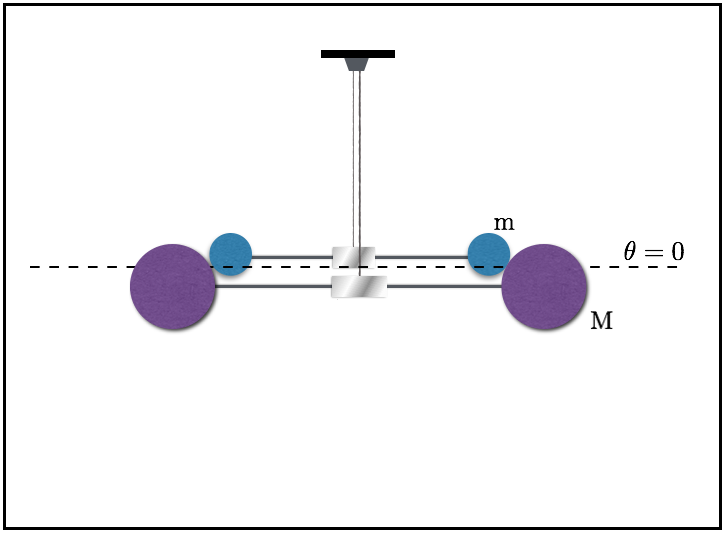}
        \caption{front-view of the oscillators}
        \label{fig:p2}
    \end{subfigure}
    \captionsetup{justification=centering}
\caption{The proposed set-up consists of two freely-moving angular oscillators suspended with vertical displacement $h$ between them, and moving angularly in the horizontal plane to which they are fixed. At the centre of each oscillator is a mirror that forms the oscillating part of a cavity system, whose other part is a fixed mirror at distance $d$ away as in (b). A focusing lens is used to reduce leakage of cavity photons due to reflections from angularly oscillating mirrors. Photons with high radiation pressure are put in a superposition of either entering the cavity or not, with a phase-mixer added to compensate for the optical path-length difference. Photons from the two paths in each cavity system are allowed to interfere independently from the photons in the other cavity before detection and analysis, as in (a).}
\label{fig:1} 
\end{figure*}
Fig. 1 shows the experimental set-up. Two micro-rods of length $2L$ each are suspended from their centre with a relative vertical separation $h$. Masses of mass $M$ and $m$ respectively are fixed at the ends of each rod and mirrors are attached to the centre of each of the rods. The mirrors will form the end mirrors, which act as mechanical oscillators, of high-finesse optical cavities. We will make use of the scheme introduced in \cite{Kleckner2008} which can place the movable end mirror of a cavity in a superposition through interaction with superposed cavity modes.

The cavity modes are generated using two high-radiation pressure photon sources, one for each cavity. Before entering the cavities, the path of each input pulse is  split using a beam splitter into two paths, one going into the cavity while the other simply passes through a phase mixer to compensate for the optical path-length difference.  A lens will be placed in the cavity to focus the incoming beam onto an edge of the mirror, so that the time needed to cross the length $d$ of the empty cavity is much smaller than the period of the rod. This set-up will generate two sets of interference patterns -- one for each rod-- between the two optical paths. 

The Hamiltonian describing the interaction between the cavity modes with the mirrors is given by \cite{Law1995}
\begin{align}
\label{eq: OldHamiltonian}
H_1 &= \hbar\omega_c(c_1^{\dagger}c_1+c_2^{\dagger}c_2)+\hbar\Omega_aa^{\dagger}a-\Lambda_m\hbar\Omega_ac_1^{\dagger}c_1(a^{\dagger}+a)\nonumber\\
& +\hbar\omega_d(d_1^{\dagger}d_1+d_2^{\dagger}d_2)+\hbar\Omega_bb^{\dagger}b-\Lambda_M\hbar\Omega_bd_1^{\dagger}d_1(b^{\dagger}+b)
\end{align}
where $a$ and $a^{\dagger}$ (respectively $b$ and $b^{\dagger}$) are the creation and annihilation operators of the mechanical modes of rod $m$ ($M$),  $c_1$ and $c_1^{\dagger}$ ($d_1$ and $d_1^{\dagger}$) are the creation and annihilation operators of photons in the path entering the cavity containing the mirror attached on rod $m$ ($M$) while $c_2$ and $c_2^{\dagger}$ ($d_2$ and $d_2^{\dagger}$) are those of photons in the path not entering the cavity. In addition, $\omega_c$ and $\omega_d$ are the frequencies of the two input pulses, $\Omega_a$ and $\Omega_b$ are the natural frequencies of the two rods of masses $m$ and $M$, respectively, and 
 \begin{align}
 \Lambda_m &= \frac{\omega_c}{2d\; \Omega_a}\sqrt{\frac{\hbar}{m\Omega_a}}\qquad 
  \Lambda_M = \frac{\omega_d}{2d\; \Omega_b}\sqrt{\frac{\hbar}{M\Omega_b}}
  \end{align}
  are the optomechanical coupling constants \cite{Kleckner2008}. The rods are assumed initially to be in coherent oscillatory states 
\begin{equation}
\ket{\beta_j} = \sum_{n=0}^{\infty} \frac{\beta_j^n}{\sqrt{n!}}\ket{n},\quad j\in\{m,M\}
\end{equation}
where $\ket{n}$ are the Fock eigenstates of the harmonic oscillator. The initial state of the total system is
\begin{align}
\label{eq: Initial State}
\ket{\psi(0)} &= \frac{1}{\sqrt{2}}\left(\ket{0,1}_c+\ket{1,0}_c\right)\ket{\beta_m}\nonumber\\
&\qquad \otimes  \frac{1}{\sqrt{2}}\left(\ket{0,1}_d+\ket{1,0}_d\right)\ket{\beta_M}
\end{align}
where $\ket{1,0}_{\chi} = \chi_1^{\dagger}\ket{0}$, $\ket{0,1}_{\chi} = \chi_2^{\dagger}\ket{0}$ for $\chi = c,d$ and where $\ket{0}$ is the vacuum state of the cavity modes. Under the action of $H_1$, this state evolves to \cite{Mancini1997}
\begin{align}
&\ket{\psi(t)} = e^{-iH_1t}\ket{\psi(0)}\\
&= \frac{e^{-i\omega_ct}}{\sqrt{2}}\left(\ket{0,1}_c\ket{\Phi_{0,m}(t)}+e^{i\phi_m(t)}\ket{1,0}_c\ket{\Phi_{1,m}(t)}\right)
\nonumber\\
&\quad \otimes \frac{e^{-i\omega_dt}}{\sqrt{2}}\left(\ket{0,1}_d\ket{\Phi_{0,M}(t)}+e^{i\phi_M(t)}\ket{1,0}_d\ket{\Phi_{1,M}(t)}\right)
\nonumber
\end{align}
where
\begin{align}
\Phi_{0,j}(t) &= \beta_j e^{-i\Omega_kt} \nonumber\\
\Phi_{1,j}(t) &= \beta_j e^{-i\Omega_kt}+\Lambda_j(1-e^{-i\Omega_kt}) \\
\phi_j(t) & = \Lambda_j^2(\Omega_kt-\sin\Omega_kt)+\Lambda_j\Im[\beta_j(1-e^{-i\Omega_kt})]
 \nonumber
\end{align}
for $(j,k)\in\{(m,a),(M,b)\}$. The interferometric visibility pattern is directly measurable from the statistics of photon detection and, therefore, it provides an important source of information about the cavity system. If the system evolves only according to $H_1$, then the visibility pattern on the photons of the two cavities will be
\begin{align}
\mathcal{V}_{0,c}(t) &= e^{-\Lambda_m^2(1-\cos\Omega_at)}\\
\mathcal{V}_{0,d}(t) &= e^{-\Lambda_M^2(1-\cos\Omega_bt)}\nonumber
\end{align}
which shows the independence of each cavity system from the other, and that the timescale of oscillation of the visibility pattern is set by the frequency of the oscillating rod. In this case, the visibility is given by twice the absolute value of one of the off-diagonal terms in the photon density matrix so that, for instance, if $\rho_c$ is the reduced density matrix of the photon coupled to rod $m$ then $\mathcal{V}_{0,c}(t) = 2\left|\Tr[\rho_{0,c}(t)\ket{0,1}_c\bra{1,0}_c]\right|$. 

Our set-up is designed so as to maximize the effect of the gravitational interaction between the two oscillators. Assuming Newtonian gravity, the total quantized Hamiltonian of the system of interacting oscillators, up to a constant term, is (see appendix A)
\begin{align}
\label{eq: NewHamiltonian}
H &= \hbar\omega_c(c_1^{\dagger}c_1+c_2^{\dagger}c_2)+\hbar\omega_aa^{\dagger}a-\lambda_m\hbar\omega_ac_1^{\dagger}c_1(a^{\dagger}+a)\nonumber\\
&+ \hbar\omega_d(d_1^{\dagger}d_1+d_2^{\dagger}d_2)+\hbar\omega_bb^{\dagger}b-\lambda_M\hbar\omega_bd_1^{\dagger}d_1(b^{\dagger}+b)\nonumber\\
&+ \hbar\gamma (a^{\dagger}+a)(b^{\dagger}+b).
\end{align}
We will denote
\begin{equation}
H_g := \hbar\gamma(a^{\dagger}+a)(b^{\dagger}+b),
\end{equation}
where
\begin{equation}
\gamma := -\frac{G}{2h^3}\sqrt{\frac{Mm}{\omega_a\omega_b}}
\end{equation}
is the gravitational coupling constant between the two oscillators. We note also that the frequencies of the oscillators and the optomechanical coupling constant is modified from the old Hamiltonian in Eq. \eqref{eq: OldHamiltonian} according to
\begin{align}
\Omega_a &\rightarrow\omega_a = \sqrt{\Omega_a^2+\frac{GM}{h^3}}, &\quad \Omega_b &\rightarrow\omega_b = \sqrt{\Omega_b^2+\frac{Gm}{h^3}}\\
\Lambda_m &\rightarrow\lambda_m = \frac{\omega_c}{2\omega_ad}\sqrt{\frac{\hbar}{m\omega_a}}, &\quad \Lambda_M &\rightarrow\lambda_M = \frac{\omega_d}{2\omega_bd}\sqrt{\frac{\hbar}{M\omega_b}}
\end{align}
The visibility pattern of photons in the coupled system will be different from that of the uncoupled system given in Eq. (8). To calculate this shift, we switch to the interaction picture in which the density matrix of the total system is
\begin{equation}
\rho_I(t) = U(t)\rho_I(0)U^{\dagger}(t)
\end{equation}
where $\rho_I(0) = \ket{\psi(0)}\bra{\psi(0)}$, 
\begin{equation}
U(t) = \mathcal{T} \exp\left[-\frac{i}{\hbar}\int_0^tdt'H_I(t')\right]
\end{equation}
with $\mathcal{T}$ being the time-ordering operator, and 
\begin{align}
H_I(t) &= e^{iH_0t/\hbar}H_ge^{-iH_0t/\hbar} \\
&= \hbar\gamma(a^{\dagger}e^{i\omega_at}+ae^{-i\omega_at}+2\lambda_mc_1^{\dagger}c_1(1-\cos\omega_at))\nonumber\\
&\times (b^{\dagger}e^{i\omega_bt}+be^{-i\omega_bt}+2\lambda_Md_1^{\dagger}d_1(1-\cos\omega_bt))
\nonumber
\end{align}
(see appendix B) where $H_0$ is comprised of the first two lines of Eq. \eqref{eq: NewHamiltonian}. The expectation value of any operator $\mathcal{O}$ is independent of the picture used to calculate it. In the interaction picture, this is equal to
\begin{align}
\braket{\mathcal{O}(t)} &= \Tr[\rho_I(t)\mathcal{O}_I(t)]\nonumber\\
&= \Tr[U(t)\rho_I(0)U^{\dagger}(t)e^{iH_0 t/\hbar}\mathcal{O}_Se^{-iH_0 t/\hbar}]\nonumber\\
&= \Tr[e^{-iH_0 t/\hbar}U(t)\rho_I(0)U^{\dagger}(t)e^{iH_0 t/\hbar}\mathcal{O}_S]
\end{align}
where $\mathcal{O}_S$ is the operator in the Schrodinger picture. 
 \begin{figure*}[htb]
 \centering
    \begin{subfigure}{0.45\textwidth}
        \includegraphics[width=1.0\linewidth]{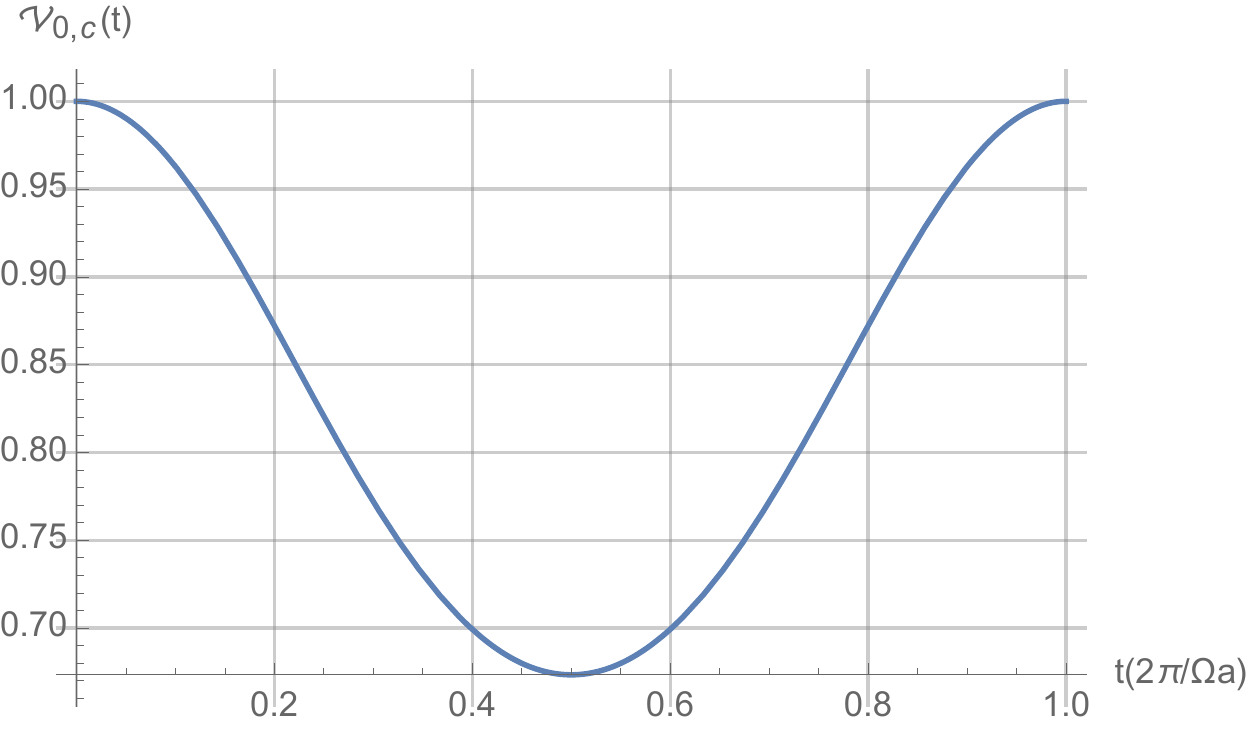}
        \caption{}
        \label{fig:p1}
    \end{subfigure}
    \begin{subfigure}{0.47\textwidth}
        \includegraphics[width=1.1\linewidth]{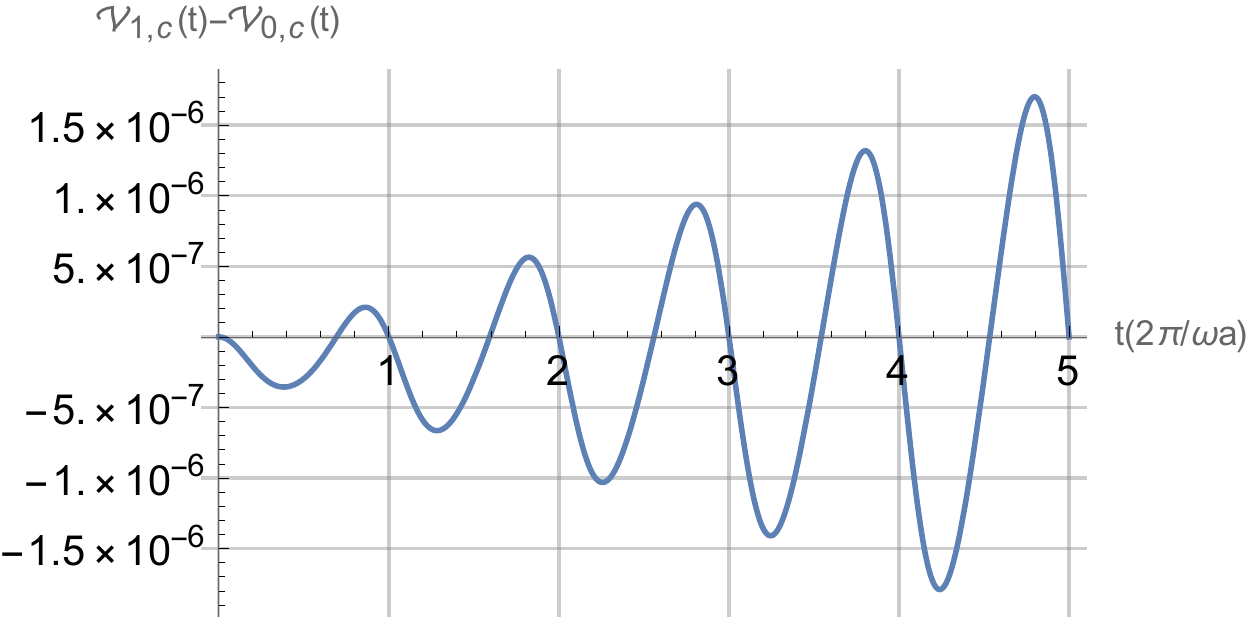}
        \caption{}
        \label{fig:p2}
    \end{subfigure}
        \captionsetup{justification=centering}
\caption{(a) The visibility pattern of the photon field in the cavity system of rod $m$ before coupling it to rod $M$, showing periodic behaviour whose period is determined by that of the oscillator $2\pi/\Omega_a$ and the strength of its drop at every half-period depends on the optomechanical coupling between the rod and the photon field. (b) The shift in the magnitude of visibility from the case with no gravitational coupling as a function of time due to the combined effect of the modified frequency of the oscillator, $\Omega_a\rightarrow\omega_a$, and the action of the coupled Hamiltonian on the state of the system and calculated perturbatively in Eq (19).}
\label{fig:1} 
\end{figure*}
The visibility of photons in the cavity of rod $m$ due to the full Hamiltonian is therefore
\begin{equation}\label{vis1}
\mathcal{V}_{1,c}(t) = 2\left|\Tr[\rho_{1,c}(t)\ket{0,1}_c\bra{1,0}_c]\right|
\end{equation}
where
\begin{equation}
\rho_{1,c}(t) =\Tr_{m,M,d}\left[e^{-iH_0 t/\hbar}U(t)\rho_I(0)U^{\dagger}(t)e^{iH_0 t/\hbar}\right]
\end{equation}
is the partial state of the photons in the cavity of rod $m$ in the Schrodinger picture after tracing out the two oscillators and the photons in the cavity of rod $M$.  The visibility \eqref{vis1} of photons in the cavity of rod $m$ is
\begin{widetext}
\begin{align}
\label{eq: VisibilityShift}
\mathcal{V}_{1,c}(t) &\approx e^{-\lambda_m^2(1-\cos\omega_at)}\times \left|1+i2\gamma\int_0^tdt'\lambda_m(1-\cos\omega_a(t'-t))\times (2\beta_M\cos\omega_bt'+\lambda_M(1-\cos\omega_bt'))\right|\\
&= e^{-\lambda_m^2(1-\cos\omega_at)}\times \left|1+i2\gamma\lambda_m[(2\beta_M-\lambda_M)(\frac{\sin\omega_bt}{\omega_b}-\frac{\omega_a\sin\omega_at-\omega_b\sin\omega_bt}{\omega_a^2-\omega_b^2})+\lambda_M(t-\frac{\sin\omega_at}{\omega_a})]\right|\nonumber
 \end{align}
 \label{eq: Visibility}
 \end{widetext}
 to first order in $\gamma$ (see appendix C). 
 
Quantum optomechanics allows coherent quantum control over massive mechanical objects ranging from nano-sized devices of $10^{-20}$ kg, to micro-mechanical structures of masses $10^{-11}$ kg, up to centimeter-sized suspended mirrors of several kilograms in mass for gravitational wave detectors \cite{Aspelmeyer2014}. We assume the masses attached to the end of the rods to be micro-mechanical structures with masses $M=m=10^{-13}$ kg and to be separated by a vertical distance $h = 10^{-8}$ m, each mounted on an oscillator with frequencies $\Omega_a = 3\times 10^3$ Hz, and $\Omega_b = \alpha\Omega_a$ for $\alpha = 0.9$. The oscillators are assumed to be cooled down to near their ground states so that $\beta_M=\beta_m=1$. We propose to use light of frequency $\omega_c=\omega_d=450\times 10^{12}$ Hz in both cavities, each with cavity length $d = 10$ cm. 
 
 The resulting initial visibility pattern, $\mathcal{V}_{0,c}(t)$, and the shift in visibility induced by the gravitational interaction, $\mathcal{V}_{1,c}(t)-\mathcal{V}_{0,c}(t)$, for photons of the cavity system of $m$ are both shown in Fig. 2. In Fig. 2(a), we see that the visibility pattern of cavity photons in the non-interacting system has the same period $2\pi/\Omega_a$ as the oscillator, and at half   that period it reaches its minimum point at $e^{-2\Lambda_m^2}$. The drop in visibility  in the middle of the period is because oscillations of the rod contain which-path information about the position of the superposed photons, dependent on the coupling strength $\lambda_m$ between the photon field and the oscillator. When the oscillator returns to its original position after a full period of oscillation, this which-path information is deleted and the visibility is restored  to its original value.

Fig. 2(b) shows the shift in visibility as a function of time, when the two oscillators are coupled to each other via Newtonian gravity. The sources of this shift are twofold. 
The first  is due to the difference in frequencies between the coupled oscillators and their idealized uncoupled state. This  is observable as a shift in the frequency of the visibility pattern of photons of magnitude $\omega_k-\Omega_k\approx\frac{GMm}{2jh^3\Omega_k}\sim \mathcal{O}(\gamma)$, for $(j,k)\in\{(m,a),(M,b)\}$. The second kind of shift is due to the second term in Eq. (19), which oscillates around ${(\gamma\lambda_m\lambda_M)}^2t^2\sim\mathcal{O}(\gamma^2)$ and is observable as a growing variation in the shape of the visibility pattern from the one in $e^{-\lambda_m^2(1-\cos\omega_at)}$. Recall also that $\lambda_M$ is the coupling parameter between the mirror in the cavity of rod $M$ and its cavity mode. From Eq. \eqref{eq: VisibilityShift}, we see that when this coupling is turned off ($\lambda_M =0$), the shift in visibility is still   that of Fig. 2(b) for small times. However  the effect of the coupling is an increase in the shift with time due to the $\lambda_M(t-\frac{\sin(\omega_at)}{\omega_a})$ term. Maintaining the coherence of the state for longer times will therefore lead to more observable effects.

\section{Discussion}

Reminiscent of the experiment done by Cavendish \cite{Cavendish} using suspended masses to measure the gravitational interaction between them, a Quantum Cavendish Experiment is   one that uses suspended masses in a quantized state to detect and measure gravitational effects in quantum regimes so that the effects of Earth's gravity  cancel out.
 Such types of experiments have been used before in sensitive verification of Newton's inverse-square law at scales below the dark-energy length scale \cite{Kapner2007}, and have been first incorporated in an optomechanical set-up to approach the quantum limit of mechanical sensing in \cite{Kim2016}. Recent proposals have considered its application in testing gravitational decoherence models \cite{Carlesso2017}, and its implementation using optically levitated nano-dumbbells \cite{Ahn2018}. Quantization of suspended linearly moving mirrors whose dynamics is dominated by the radiation-pressure of cavity photons has been achieved with masses ranging from $40$ kg \cite{Harry2010} to milli-grams \cite{Matsumoto2015}. 
 
 Our set-up requires forming coherent states of torsional mirrors of nanogram masses by cooling them to their ground states,  surpassing the standard quantum limit of detection \cite{Enomoto2016}. The suspended masses are coupled to a cavity field inside an optomechanical set-up, and the effect of the mutual gravitational interaction between the masses is calculated on the visibility pattern of cavity photons, whose observation is based on robust and well-tested experimental techniques. 
 
 We found that the effects on the visibility are of two types: a shift in the period of revival by an amount $\delta T=\frac{2\pi}{\Omega_a}-\frac{2\pi}{\omega_a}$, and a change in the shape of the visibility pattern from the functional form $e^{-\lambda_m^2(1-\cos\omega_at)}$ that is of order $\mathcal{O}(\gamma^2)$ for time scale $t\lesssim\gamma^{-1}$. In practice, it is easier to detect $\delta T$, which corresponds to $\delta T\approx 0.78$ ns for the parameters used above, than the shift in vertical magnitude that is of order $10^{-6}$ in Fig. 2(b). 
 
 To illustrate, suppose that the visibility at some time $t$ is drawn from an a priori Gaussian distribution of variance $\sigma^2$. Then the error on the estimate of the visibility at time $t$ obtained by averaging over $N$ data points is $\sigma_{\text{error}}=\frac{\sigma}{\sqrt{N}}$. If $\sigma_{\text{error}}\sim 10^{-6}$ then $N\sim 10^{12}\sigma^2 $, which is difficult to achieve. On the other hand, the accuracy of measuring $\delta T$ is dependent only on the time resolution available.

In practice an oscillator in a coherent state $\ket{\beta}\bra{\beta}$ will be in a thermal mixture 
\begin{equation}
\frac{1}{\pi\bar{n}}\int d^2\beta e^{-{|\beta|}^2/\bar{n}}\ket{\beta}\bra{\beta}
\end{equation}
where $\bar{n}={(e^{\hbar\omega_a/k_BT}-1)}^{-1}$ is the mean thermal number of phonon excitations at temperature $T$. This will modify the visibility according to \cite{Marshall2003}
\begin{equation}
e^{-\lambda_m^2(1-\cos\omega_at)} \rightarrow e^{-\lambda_m^2(2\bar{n}+1)(1-\cos\omega_at)}
\end{equation}
which causes a fast decay in visibility that is revived only after a full period. The width of the visibility's revived peak scales according to $\sim\frac{1}{\lambda_m\sqrt{\frac{4k_BT}{\hbar\omega_a}+2}}$. Increasing this width constitutes one of the main experimental challenges to realize this proposal, and requires a method to cool down the centre of mass mode of oscillator to very near their ground state \cite{Bhattacharya2007}. Another experimental challenge is due to decoherence from the mechanical damping of the oscillator and from dephasing with the environment, which lowers the revived peak of visibility. If the environment is modelled as an Ohmic thermal bath of harmonic oscillators and the damping rate of oscillators is $\Gamma_a$, then the dephasing rate due to the environment at temperature $T$ is $\Gamma_D=\Gamma_ak_BTm{(\Delta x)}^2/\hbar^2$, where $\Delta x \sim \sqrt{\frac{\hbar}{m\omega_a}}$ is the uncertainty in position of the oscillator \cite{Zeh2003}. The condition for environmental decoherence is then $\Gamma_D\lesssim\omega_a$, which corresponds to 
\begin{equation}
Q\gtrsim\frac{k_BT}{\hbar\omega_a}\sim\bar{n}
\end{equation}
where $Q:=\omega_a/\Gamma_a$ is the quality factor of the oscillator. Values of $Q\sim 10^7$ have been achieved for suspended nanoparticles \cite{Gieseler2012}, which corresponds  to $T\lesssim 0.23$ K for the parameters of the set-up considered here.

A novel feature of our proposed scheme is the observation of effects arising from gravitationally interacting quantum systems (whereas most previous studies are for a quantum test mass in the background gravitational field of the Earth). It is also interesting to note that entanglement, albeit quite weak,  is generated due to this gravitational interaction. Denoting for convenience the system associated with $m$ system $1$ (consisting of the oscillating mirror and the cavity modes) and that of $M$ as system $2$, we see that the initial state $\ket{\psi(0)}\bra{\psi(0)}$ in Eq. \eqref{eq: Initial State} is separable between the two systems. Since the only coupling between systems $1$ and $2$ in the proposed scheme is gravity, any resulting entanglement between the two systems can be attributed to the gravitational force. 
\begin{figure}
    \centering
    \includegraphics[scale=0.6]{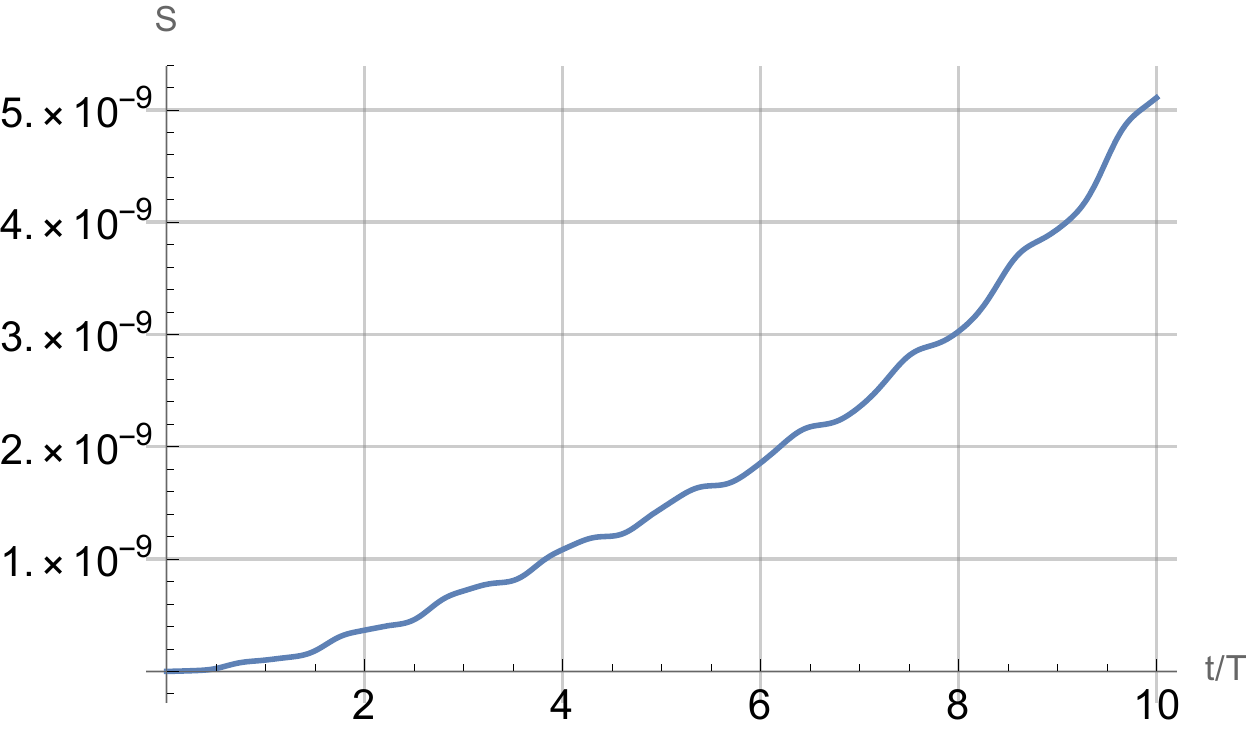}
    \caption{Plot of Linear Entropy, $S$, against $\frac{2\pi }{\omega_a}t$.}
    \label{fig: LinearEntropy}
\end{figure}
To quantify the amount of entanglement, we can use the linear entropy, defined as $S = 1- \Tr (\rho_1^2)$, where $\rho_1$ is the partial state of system $1$. The calculation of linear entropy is carried out using the same set of parameters above, and is given in appendix D. Fig. \ref{fig: LinearEntropy} shows the result. Even though the amount of entanglement generated for the period shown is small, we observe an increase with time, similar to the shift in the visibility pattern. Since the visibility is related to how much which-path information the position of the oscillator can reveal, which in turn is dependent on the amount of entanglement between the oscillator and the cavity photons, the increase in the amount of entanglement due to gravity shown in Fig. 3 means that, by monogamy of entanglement, the correlations between the oscillator and the cavity photons will correspondingly decay. This causes the visibility pattern to have a growth term as given in Eq. (19). 
We expect that an exact calculation will give a linear entropy and visibility that are bounded from above. Given the recent interest in observing entanglement due to gravity \cite{Marletto2017,Bose2017}, it will be desirable to obtain an entanglement witness that can experimentally verify the entanglement generated for this scheme.

\section*{Acknowledgements}
This work was supported in part by the Natural Sciences Engineering Research Council. We thank Dan Carney for useful comments.
\begin{appendices}
\section{Total Hamiltonian with Newtonian gravity}
Two classical harmonic oscillators will couple via gravity according to
\begin{equation}
H = \frac{p_m^2}{2m}+\frac{1}{2}I_m\Omega_a^2\theta_m^2+\frac{p_M^2}{2M}+\frac{1}{2}I_M\Omega_b^2\theta_M^2+H_g
\end{equation}
where $I_m = 2mL^2$ and $I_M = 2ML^2$ are the two moments of inertia for the two rods. For two angular oscillators with masses at each end of a rod of length $L$ and suspended with vertical displacement $h$, the classical gravitational interaction will be
\begin{align}
H_g & = 2\frac{-GMm}{{(h^2+{(2L\sin(\frac{\theta_M-\theta_m}{2}))}^2)}^{1/2}}\\
& \approx 2\frac{-GMm}{h{(1+{(L(\theta_M-\theta_m)/h)}^2)}^{1/2}}\nonumber\\
& \approx  2\frac{-GMm}{h}\left(1-\frac{1}{2}{(\frac{L(\theta_M-\theta_m)}{h})}^2\right)\nonumber\\
& = \frac{-2GMm}{h}+\frac{GMm}{h^3}{(L(\theta_M-\theta_m))}^2\nonumber\\
& = \frac{-2GMm}{h}+\frac{GMmL^2}{h^3}\left(\theta_M^2+\theta_m^2-\theta_M\theta_m-\theta_m\theta_M\right)\nonumber
\end{align}
Therefore, up to a constant term, the total Hamiltonian can be written as
\begin{equation}
H = \frac{p_m^2}{2m}+\frac{1}{2}m\omega_a^2\theta_m^2+\frac{p_M^2}{2M}+\frac{1}{2}M\omega_b^2\theta_M^2-\frac{2GMmL^2}{h^3}\theta_m\theta_M
\end{equation}
where $\omega_a = \sqrt{\Omega_a^2+\frac{GM}{h^3}}$ and $\omega_b = \sqrt{\Omega_b^2+\frac{Gm}{h^3}}$. The frequency of a photon inside a cavity of length $d$ is
\begin{equation}
\omega_c = 2\pi \frac{nc}{2d} = \frac{n\pi c}{d}
\end{equation}
where $n=1,2,3,\dots$ and $c$ is the speed of light. When it couples to an angular oscillator with displacement $\theta$, the length of the cavity varies $d\rightarrow d+\delta$ for $\delta<<d$, so that
\begin{equation}
\omega_c = \frac{n\pi c}{d+\delta} = \frac{n\pi c}{d(1+\delta/d)} \approx \frac{n\pi c}{d}(1-\delta/d)
\end{equation}
In our case, $\delta = L\sin\theta \approx L\theta$. So, 
\begin{equation}
\omega_c \rightarrow \omega_c-\omega_c \frac{L\theta}{d}
\end{equation}
Introducing the annihilation operators for the two oscillators
\begin{align}
a &= \sqrt{\frac{I_m\omega_a}{2\hbar}}\left(\theta_m+\frac{i}{I_m\omega_a}p_m\right)\\
b &= \sqrt{\frac{I_M\omega_b}{2\hbar}}\left(\theta_M+\frac{i}{I_M\omega_b}p_M\right)
\end{align}
and substituting back in the total Hamiltonian to rewrite it in terms of the creation/annihilation operators, including the photon cavity terms gives the quantized Hamiltonian of the total system as
\begin{align}
H &= \hbar\omega_c(c_1^{\dagger}c_1+c_2^{\dagger}c_2)+\hbar\omega_aa^{\dagger}a-\lambda_m\hbar\omega_ac_1^{\dagger}c_1(a^{\dagger}+a)\nonumber\\
&+ \hbar\omega_d(d_1^{\dagger}d_1+d_2^{\dagger}d_2)+\hbar\omega_bb^{\dagger}b-\lambda_M\hbar\omega_bd_1^{\dagger}d_1(b^{\dagger}+b)\nonumber\\
&+ \hbar\gamma (a^{\dagger}+a)(b^{\dagger}+b)
\end{align}
exactly as given in Eq.9.
\section{Interaction Hamiltonian}
We will derive here the expression in Eq.16 for the interaction Hamiltonian $H_I(t) = e^{iH_0t/\hbar}H_ge^{-iH_0t/\hbar}$. Given operators $A$ and $B$, the BHC formula is
\begin{equation}
e^ABe^{-A} = B+[A,B],\frac{1}{2}[A,[A,B]]+\dots
\end{equation}
The operator $e^{-iH_0t/\hbar}$ was calculated to be \cite{Bose1997}
\begin{align}
e^{-iH_0t/\hbar} &= e^{-i\omega_c t(c_1^+c_1+c_2^+c_2)}e^{i{(\lambda_mc_1^+c_1)}^2(\omega_at-\sin(\omega_at))} \nonumber\\
& e^{\lambda_mc_1^+c_1\left(a^+\alpha-a\alpha^*\right)}e^{-i\omega_ata^+a}\times [M]
\end{align}
where $\alpha = (1-e^{-i\omega_at})$, and $[M]$ here and below denotes the same part of the term as on its left but under the isomorphic transformations
\begin{align*}
{(.)}_{a,c,m} &\rightarrow {(.)}_{b,d,M}\\
a,c &\rightarrow b,d
\end{align*}
Using the BHC formula, the interaction Hamiltonian can be written as
\begin{align}
H_I(t) &= e^{iH_0t/\hbar}\hbar\gamma(a^{\dagger}+a)(b^{\dagger}+b)e^{-iH_0t/\hbar}\\
 &= \hbar\gamma e^{i\omega_ata^{\dagger}a}e^{\lambda_mc_1^{\dagger}c_1(a\alpha^*-a^{\dagger}\alpha)} (a^{\dagger}+a) \nonumber\\
 & e^{-\lambda_mc_1^{\dagger}c_1(a\alpha^*-a^{\dagger}\alpha)}e^{-i\omega_ata^{\dagger}a}\times[M]\nonumber\\
&= \hbar\gamma e^{i\omega_ata^{\dagger}a}(a^{\dagger}+a+\lambda_mc_1^{\dagger}c_1(\alpha+\alpha^*))e^{-i\omega_ata^{\dagger}a}\times[M]\nonumber\\
&=\hbar \gamma (a^{\dagger}e^{i\omega_at}+ae^{-i\omega_at}+\lambda_mc_1^{\dagger}c_1(\alpha+\alpha^*))\times [M]\nonumber\\
&= \hbar\gamma (a^{\dagger}e^{i\omega_at}+ae^{-i\omega_at}+2\lambda_mc_1^{\dagger}c_1(1-\cos\omega_at))\times [M]\nonumber
\end{align}
which is Eq.16.
\section{Visibility in the coupled system}
To calculate the visibility from Eq.18, we need to know what the action of $e^{-iH_0t\hbar}U(t)$ on $\rho_I(0)$ is. First, note that
\begin{align}
e^{-iH_0t/\hbar}a^{\dagger} &= e^{-iH_0t/\hbar}a^{\dagger}e^{iH_0t/\hbar}e^{-iH_0t/\hbar}\nonumber\\
&= e^{-\lambda_mc_1^{\dagger}c_1(a\alpha^*-a^{\dagger}\alpha)}e^{-i\omega_ata^{\dagger}a}a^{\dagger}e^{i\omega_ata^{\dagger}a}\nonumber\\
& e^{\lambda_mc_1^{\dagger}c_1(a\alpha^*-a^{\dagger}\alpha)}e^{-iH_0t/\hbar}\nonumber
\end{align}
\begin{align}
&= e^{-\lambda_mc_1^{\dagger}c_1(a\alpha^*-a^{\dagger}\alpha)}a^{\dagger}e^{-i\omega_at}e^{\lambda_mc_1^{\dagger}c_1(a\alpha^*-a^{\dagger}\alpha)}e^{-iH_0t/\hbar}\nonumber\\
&= (a^{\dagger}-\lambda_mc_1^{\dagger}c_1\alpha^*)e^{-i\omega_at}e^{-iH_0t/\hbar}
\end{align}
and, similarly
\begin{align}
e^{-iH_0t/\hbar}a &= (a-\lambda_mc_1^{\dagger}c_1\alpha)e^{i\omega_at}e^{-iH_0t/\hbar}
\end{align}
This allows us to write, using the BHC formula, up to first-order
\begin{widetext}
\begin{align}
e^{-iH_0t/\hbar}U(t) &\approx e^{-iH_0t/\hbar}\left(1-\frac{i}{\hbar}\int_0^tdt'H_I(t')\right)\nonumber\\
&= e^{-iH_0t/\hbar}\left(1-i\gamma\int_0^tdt' (a^{\dagger}e^{i\omega_at'}+ae^{-i\omega_at'}+2\lambda_mc_1^{\dagger}c_1(1-\cos\omega_at'))\times [M]\right)\nonumber\\
&= e^{-iH_0t/\hbar}-\frac{i\gamma}{\hbar}\int_0^tdt'((a^{\dagger}-\lambda_mc_1^{\dagger}c_1\alpha^*)e^{i\omega_a(t'-t)}+(a-\lambda_mc_1^{\dagger}c_1\alpha)e^{-i\omega_a(t'-t)}\nonumber\\
&+2\lambda_mc_1^{\dagger}c_1(1-\cos\omega_at'))\times[M]e^{-iH_0t/\hbar}\nonumber\\
&= 1-\frac{i\gamma}{\hbar}\int_0^tdt'(a^{\dagger}e^{i\omega_a(t'-t)}+ae^{-i\omega_a(t'-t)}+\lambda_mc_1^{\dagger}c_1(2-2\cos\omega_at'-\alpha^* e^{i\omega_a(t'-t)}-\alpha e^{-i\omega_a(t'-t)}))\times [M]e^{-iH_0t/\hbar}\nonumber\\
&= 1-\frac{i\gamma}{\hbar}\int_0^tdt'(a^{\dagger}e^{i\omega_a(t'-t)}+ae^{-i\omega_a(t'-t)}+\lambda_mc_1^{\dagger}c_1(2-2\cos\omega_at'-2\cos\omega_a(t'-t)+2\cos\omega_at'))\times [M]e^{-iH_0t/\hbar}\nonumber\\
&= 1-\frac{i\gamma}{\hbar}\int_0^tdt'(a^{\dagger}e^{i\omega_a(t'-t)}+ae^{-i\omega_a(t'-t)}+2\lambda_mc_1^{\dagger}c_1(1-\cos\omega_a(t'-t)))\times [M]e^{-iH_0t/\hbar}
\end{align}
Using this relation, we can calculate the action of this operator on the initial state ${\ket{\psi(0)}}_I$ of the total system given in Eq.5. perturbatively to be
\begin{align}
e^{-iH_0t/\hbar}U(t)\ket{\psi(0)}_I &= \ket{\psi(t)}-\frac{i\gamma}{\hbar}\int_0^tdt'(a^{\dagger}e^{i\omega_a(t'-t)}+ae^{-i\omega_a(t'-t)}\nonumber\\
&+2\lambda_mc_1^{\dagger}c_1(1-\cos\omega_a(t'-t)))\times [M]\ket{\psi(t)}\nonumber \\
&= \ket{\psi(t)}-\frac{i\gamma}{\hbar}\int_0^tdt'(a^{\dagger}e^{i\omega_a(t'-t)}+ae^{-i\omega_a(t'-t)}+2\lambda_mc_1^{\dagger}c_1(1-\cos\omega_a(t'-t)))\nonumber \\
& (b^{\dagger}e^{i\omega_b(t'-t)}+be^{-i\omega_b(t'-t)}+2\lambda_Md_1^{\dagger}d_1(1-\cos\omega_b(t'-t))) \ket{\psi(t)}\nonumber\\
&= \frac{e^{-i\omega_ct-i\omega_dt}}{2}\left[\right.\left(1-\frac{i\gamma}{\hbar}\int_0^tdt'(a^{\dagger}e^{i\omega_a(t'-t)}+\Phi_{0,m}(t)e^{-i\omega_a(t'-t)})\right.\nonumber \\
& \left.(b^{\dagger}e^{i\omega_b(t'-t)}+\Phi_{0,M}(t)e^{-i\omega_b(t'-t)})\right)\nonumber\\
& \ket{0,1}_c\ket{0,1}_d\ket{\Phi_{0,m}}\ket{\Phi_{0,M}}\nonumber\\
&+\left(1-\frac{i\gamma}{\hbar}\int_0^tdt'(a^{\dagger}e^{i\omega_a(t'-t)}+\Phi_{0,m}(t)e^{-i\omega_a(t'-t)})\right.\nonumber \\
& \left.(b^{\dagger}e^{i\omega_b(t'-t)}+\Phi_{1,M}(t)e^{-i\omega_b(t'-t)}+2\lambda_M(1-\cos\omega_b(t'-t)))\right)\nonumber\\
& e^{i\phi_M(t)}\ket{0,1}_c\ket{1,0}_d\ket{\Phi_{0,m}}\ket{\Phi_{1,M}}\nonumber
\end{align}
\begin{align}
& +\left(1-\frac{i\gamma}{\hbar}\int_0^tdt'(a^{\dagger}e^{i\omega_a(t'-t)}+\Phi_{1,m}(t)+2\lambda_m(1-\cos\omega_a(t'-t)))\right.\nonumber \\
& \left.(b^{\dagger}e^{i\omega_b(t'-t)}+\Phi_{0,M}(t)e^{-i\omega_b(t'-t)})\right) \nonumber\\
& e^{i\phi_m(t)}\ket{1,0}_c\ket{0,1}_d\ket{\Phi_{1,m}}\ket{\Phi_{0,M}}\nonumber\\
&+\left(1-\frac{i\gamma}{\hbar}\int_0^tdt'(a^{\dagger}e^{i\omega_a(t'-t)}+\Phi_{1,m}(t)e^{-i\omega_a(t'-t)+2\lambda_m(1-\cos\omega_a(t'-t))})\right.\nonumber \\
& \left.(b^{\dagger}e^{i\omega_b(t'-t)}+\Phi_{1,M}(t)e^{-i\omega_b(t'-t)}+2\lambda_M(1-\cos\omega_b(t'-t)))\right)\nonumber\\
& e^{i\phi_m(t)}e^{i\phi_M(t)}\ket{1,0}_c\ket{1,0}_d\ket{\Phi_{1,m}}\ket{\Phi_{1,M}}]
\end{align}
Tracing out the two oscillators and the photons in the cavity of rod M from the density matrix formed by this state, keeping terms only of order $\mathcal{O}(\gamma)$, and calculating twice the absolute value of one of the off-diagonal terms will give the expression $\mathcal{V}_{1,c}(t)$ in Eq. (19).
\end{widetext}
\section{linear entropy}
If we define $A:=\frac{-1}{\gamma\hbar}\int_0^tdt'e^{-iH_0t/\hbar}H_I(t')e^{iH_0t/\hbar}$, then we note that it is Hermitian, and from Eq. (35)-(36) that it can be written as $A=\frac{-1}{\gamma\hbar}\int_0^tdt'H_I(t'-t)$. The density matrix of the two systems: system 1 for oscillator of mass $m$ with its cavity photons and system 2 for oscillator of mass $M$ with its cavity photons, can be written as a separable pure bi-partite state $\rho=\rho_1\otimes\rho_2=: \ket{\psi_1}\bra{\psi_1}\otimes\ket{\psi_2}\bra{\psi_2}$ so that $\ket{\psi_1}\ket{\psi_2}=\ket{\psi(t)}$, as given in Eq. (5). Further defining
\begin{align}
{A_1}^2 &:= {(\bra{\psi_2}A\ket{\psi_2})}^2\\
{A^2}_1 &:= \bra{\psi_2}A^2\ket{\psi_2}
\end{align}
if we use the BCH formula in Eq. (32) and calculate up to second-order in $\gamma$, then under the action of the unitary $U=e^{i\gamma A}$, the state in the Schrodinger picture evolves according to
\begin{align}
\rho' &= U\rho U^{\dagger}\nonumber\\
&= e^{i\gamma A}\rho e^{-i\gamma A}\nonumber\\
&= \rho+i\gamma [A,\rho]+\frac{1}{2}[i\gamma A,[i\gamma A,\rho]]+\dots\nonumber\\
&= \rho+i\gamma [A,\rho]-\frac{\gamma^2}{2}(A^2\rho+\rho A^2-2A\rho A)
\end{align}
Tracing out system 2 will give
\begin{align}
\rho_1' &= \rho_1+i\gamma [A_1,\rho_1]-\frac{\gamma^2}{2}({A^2}_1\rho_1+\rho_1 {A^2}_1-2A_1\rho_1 A_1)
\end{align}
Squaring this and keeping terms only up to second-order in $\gamma$ will give
\begin{widetext}
\begin{align}
\rho_1'^2 &= \rho_1^2+i\gamma[A_1,\rho_1^2]-\frac{\gamma^2}{2}(2A_1\rho_1A_1\rho_1+2\rho_1A_1\rho_1A_1-2A_1\rho_1^2A_1-2\rho_1{A_1}^2\rho_1\nonumber\\
&+\rho_1{A^2}_1\rho_1+\rho_1^2{A^2}_1-2\rho_1A_1\rho_1A_1+{A^2}_1\rho_1^2+\rho_1{A^2}_1\rho_1-2A_1\rho_1A_1\rho_1)
\end{align}
\end{widetext}
Finally, taking the trace of this gives
\begin{align}
\Tr\rho_1' &= 1-\frac{\gamma^2}{2}(4\Tr({A^2}_1\rho_1)-4\Tr({A_1}^2\rho_1))\nonumber\\
&= 1-2\gamma^2(\Tr({A^2}_1\rho_1)-\Tr({A_1}^2\rho_1))
\end{align}
so that the linear entropy will now be
\begin{align}
S &:= 1-\Tr\rho_1'^2\nonumber\\
&= 2\gamma^2(\Tr({A^2}_1\rho_1)-\Tr({A_1}^2\rho_1))
\end{align}

\end{appendices}
\bibliography{ref.bib}
\end{document}